**Auroral diagnosis of solar wind interaction with Jupiter's magnetosphere**

Z. H. Yao[1,2†], B. Bonfond[2†], D. Grodent[2], E. Chané[3], W. R. Dunn[4], W. S. Kurth[5], J. E. P. Connerney[6,7], J. D. Nichols[8], B. Palmaerts[2], R. L. Guo[2], G. B. Hospodarsky[5], B. H. Mauk[9], T. Kimura[10] and S. J. Bolton[11]

[1]Key Laboratory of Earth and Planetary Physics, Institute of Geology and Geophysics, Chinese Academy of Sciences, Beijing, China

[2] Laboratoire de Physique Atmosphérique et Planétaire, STAR institute, Université de Liège, Liège, Belgium

[3]Centre for mathematical Plasma Astrophysics, KU Leuven, Celestijnenlaan, Belgium

[4]Mullard Space Science Laboratory, University College London, Dorking, UK

[5]Department of Physics and Astronomy, University of Iowa, Iowa City, IA, USA

[6]Space Research Corporation, Annapolis, MD 21403, USA.

[7]NASA Goddard Space Flight Center, Greenbelt, MD 20771, USA.

[8]Department of Physics and Astronomy, University of Leicester, Leicester, UK

[9]Applied Physics Laboratory, Johns Hopkins University, Laurel, MD, USA

[10]Frontier Research Institute for Interdisciplinary Sciences, Tohoku University, Sendai, Japan

[11]Southwest Research Institute, San Antonio, TX, USA

†Z. H. Yao and B. Bonfond contributed equally to this work
Corresponding email: z.yao@ucl.ac.uk

**Abstract**

Although mass and energy in Jupiter's magnetosphere mostly come from the innermost Galilean moon Io's volcanic activities, solar wind perturbations can play crucial roles in releasing the magnetospheric energy and powering aurorae in Jupiter's polar regions. The systematic response of aurora to solar wind compression remains poorly understood. Here we report the analysis of a set of auroral images with contemporaneous in situ magnetopause detections. We distinguish two types of auroral enhancements: a transient localized one and a long-lasting global one. We show that only the latter systematically appears under a compressed magnetopause, while the localized auroral expansion could occur during an expanded magnetopause. Moreover, we directly examine previous theories on how solar wind compressions enhance auroral emissions. Our results demonstrate that auroral morphologies can be diagnostic of solar wind conditions at planets when in situ measurements are not possible.

**Introduction**

Jupiter has the brightest aurorae of all the planets in our solar system, facilitating the remote observation of energy dissipation across vast distances[1]. The auroral power can significantly vary by orders of magnitude in time scales ranging from tens of seconds[2] to several hours[3], and can be observed at different wavelengths[4-7]. Hubble Space Telescope (HST) provides high-resolution ultraviolet (UV) images of Jupiter's aurora for more than 20 years, and have resolved key auroral components of the aurora, which consists of a main auroral oval, a dark region on the dawnside, a polar swirl region and a polar active region[8]. The main auroral oval is traditionally suggested driven by a magnetosphere-ionosphere coupling current system due to the breakdown of rigid corotation of plasma in the middle magnetosphere[9-11].

Observations of multiple light bands show auroral enhancements during solar wind compressions[4,12-15], contradicting the theoretical predictions based on steady-state assumptions[9,16]. New interpretations from time-varying modeling[17] and numerical simulations[18] were thus proposed to mitigate the proliferating conflict between observation and classical steady-state theoretical prediction. Recent study revealed that solar wind shocks and auroral brightening are coupled by very complicated relations[19]. Moreover, Kita, et al. [19] indicated that it requires substantial time for Jupiter's magnetosphere to response to solar wind shock arrival at the upstream magnetosphere boundary, i.e., the magnetopause. Juno's first 7 apojove periods provided direct examination of magnetopause compression[20], which could eliminate uncertainty in solar wind propagation models and could mostly exclude the response time to the solar wind compression at the magnetopause. Meanwhile, HST was planned to regularly monitor Jupiter's UV aurora during these orbits. Therefore, we could perform a systematic determination on the relation between the magnetopause compression and auroral activities, which is pivotal to assess the proposed interpretations from modeling and simulation investigations.

**Results**

One of the regular sequences of HST UV imaging observations in coordination with the Juno spacecraft[21] was planned from January 22 to 27 2017. Fig. 1 shows the projections of auroral images onto Jupiter's northern pole, and each image over about 40 minutes. On January 22, there was an auroral brightening around the dawn arc (Fig. 1a), which was not found in the successively available HST image in Fig. 1b (~29 hours later). Similar auroral enhancements on the dawn local times with significant expansions in latitudes have been identified as auroral dawn storms (ADS)[22]. The auroral image shown in Fig. 1c was obtained ~19 hours after Fig. 1b, which shows a global enhancement in all local times within HST's field of view. The dawn arc enhancement is relatively narrow in width, i.e., the direction perpendicular to the average main auroral oval (the white curve, indicated by the pink arrow in Fig. 1c), which is named main auroral brightening (MAB) in this letter. If we take the power in Fig. 1b (i.e., 1045 GW) as the baseline of quiet Jovian aurorae, the total auroral power in Fig. 1c is a factor of two higher than the total power in Fig. 1b. This auroral morphology remained similar and the power further increased to 2430 GW in the following HST visit (Fig. 1d, ~1.5 hour later). The thin enhanced auroral arc on the dawn to noon local times remained in Fig 1e while with significantly decreased power and return to almost quiet time auroral power in Fig 1f. Therefore, the MAB event likely lasted for about 2 to 3 days, consistent with previous reports on main auroral enhancements during solar wind compression based on the analysis of observations from HST[14] and Hisaki[23].

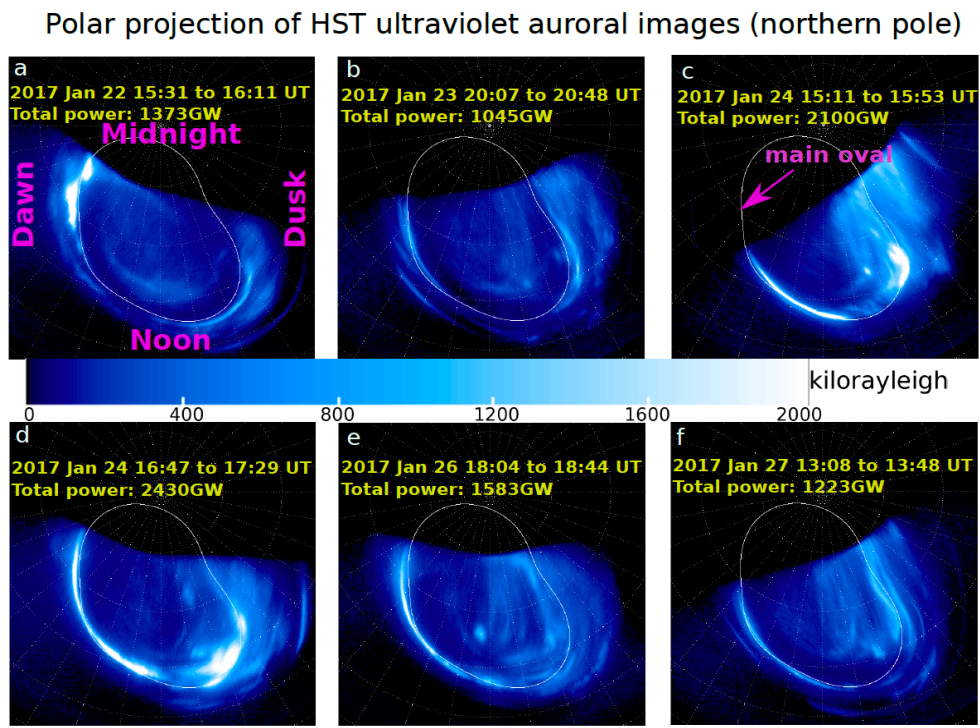

**Figure 1| Polar projections of six auroral images from January 22 to 27 2017.** Each image was averaged over ~40 minutes. The main oval (indicated by the pink

arrow in panel c) is an average main auroral oval location.

In addition to the timescales for which they exist, two different types of morphologies also allow us to distinguish the ADS in Fig. 1a from the MAB in Fig. 1c-e: (1) the MAB's enhanced dawn arc extended to near-noon local times, while the ADS in Fig. 1a was limited to the dawn local times before 9h; (2) the dawn auroral arc along the reference main oval is thinner and smoother for the MAB, but thicker and more variable along the reference oval for the ADS. We therefore define two parameters, i.e., mean arc width of the dawn aurora and the variation of auroral width along the main oval reference, for characterizing the two types of auroral morphologies. The quantitative analysis is provided in the Methods section. The mean arc width and the variation of the ADS in Fig. 1a are 1468 km and 548 km. The two values are 626 km and 262 for the MAB in Fig. 1c. The mean arc width and variation in the ADS are a factor of two larger than for the MAB. A key unsolved question is whether or not the two auroral morphologies correspond to fundamentally different drivers. Previous studies have suggested that brightening of the aurora is associated with solar wind conditions[14,18,24], however, whether solar wind conditions drive both ADS and MAB is unclear. Disentangling this solar wind influence from other drivers is critical for auroral interpretation. Since the two auroral events were successively observed separated by two days, it indicates that a complete transition between the two types of auroral morphologies could be shorter than two days. Therefore, it is insufficient to apply a modeling solar wind prediction to assess whether or not the two auroral events happened under different solar wind conditions, since the modeling prediction of solar wind condition usually involves an uncertainty of more than two days even in ideal conditions[25]. Here we directly identify magnetopause crossings using Juno's Waves instrument[26] and MAG instrument[27] in coordination with HST's auroral context. Using the magnetopause model by Joy, et al. [28], we know that the magnetopause for a compressed magnetosphere in the dawn sector is at ~90 $R_J$, whereas for an expanded magnetosphere it will at ~130 $R_J$. We can therefore identify intervals when the magnetosphere is compressed, so that we accurately assess the influences of solar wind compressions on auroral activities, and provide key information to answer two questions: (1) how does the solar wind modulate Jupiter's main aurora? (2) ADS have previously been observed during solar wind compressions[24,29], is there a physical causality or was this coincidence?

Interestingly, the hectometric radio emission with frequency of several MHz (Fig. 2a) was enhanced since January 24 when the MAB auroral event was observed, but not significantly enhanced for the ADS event on January 22. We note that the hectometric emission remained enhanced for at least two days after the MAB auroral event (e.g., Fig. 1e). Below we analyze Juno's in situ measurements to determine the extent to which solar wind conditions control the different auroral morphologies. The intense emissions with frequencies between about 200 Hz and 2 kHz (Fig. 2b) are the trapped continuum radiation[30,31]. The appearance (or disappearance) of the emission serves as a good indicator of entry into the magnetosphere (or exit into the magnetosheath)[20,32,33]. During ultraviolet auroral observations in Fig. 1, Juno traveled inbound from >110 $R_J$ to ~70 $R_J$ to the planet center (1 $R_J$ = 71,492 km) in the sector (near 05:00 Magnetic Local Time), and encountered an inward moving magnetopause on January 24 at ~78 $R_J$, so that Juno was exposed to the magnetosheath thereafter. The nominal magnetopause location on the dawnside is at > 100 $R_J$ but it can move to ~70 - 80 $R_J$ during strong compressed situations, as suggested by both models and Juno's statistical results[20,28]. On January 26th Juno returned to the magnetosphere (evidenced by the reappearance of the trapped continuum radiation), which is likely due to the recovery of magnetopause to a more probable location. Based on the wave feature, we could determine that Juno was in the magnetosheath during the period marked by the green bar on the top of panel (b) in Fig. 2, and in the magnetosphere during the rest of this period. The strongly perturbed magnetic field between January 25 and 26 also confirms that Juno was in the magnetosheath. The wave frequencies, which reflect the plasma number density[32], has significantly increased shortly before (after the first vertical dashed purple line) Juno's entry into the magnetosheath. The density increase is a naturally expected consequence of magnetopause compression[20,32,33], confirming our determination of compression from the appearance (or disappearance) of the trapped continuum radiation. Juno rapidly encountered the magnetopause after compression (marked by the first vertical dashed purple line), and the spacecraft remained in a compressed region for about two days (Jan 26-27) after re-entering into the magnetosphere. The magnetic field is highly variable only when the spacecraft was in the magnetosheath while Juno may remain in the magnetopause boundary layer for a while with distinct wave feature due to compression, so that the fluctuated magnetic field did not fully coincide with magnetopause compression associated wave feature[33]. The magnetic field components and magnetic strength from Juno (Fig. 2c-f) were nearly unperturbed before being approached by the magnetopause (as indicated by the first vertical dashed purple line), suggesting that during this period solar wind was relatively quiet. The sketch of Fig. 2g shows the motion of magnetopause together with Juno's trajectory, illustrating the strong magnetopause compression on January 24 and 25. The ADS (Fig. 1a) and quiet auroral morphology (Fig. 1b) occur during the same solar wind conditions, i.e., relatively quiet solar wind, showing that the ADS was not driven by solar wind compression. The two MAB images (Fig 1c-d) were both acquired during the compressed period (i.e., between the two vertical purple lines in Fig. 2).

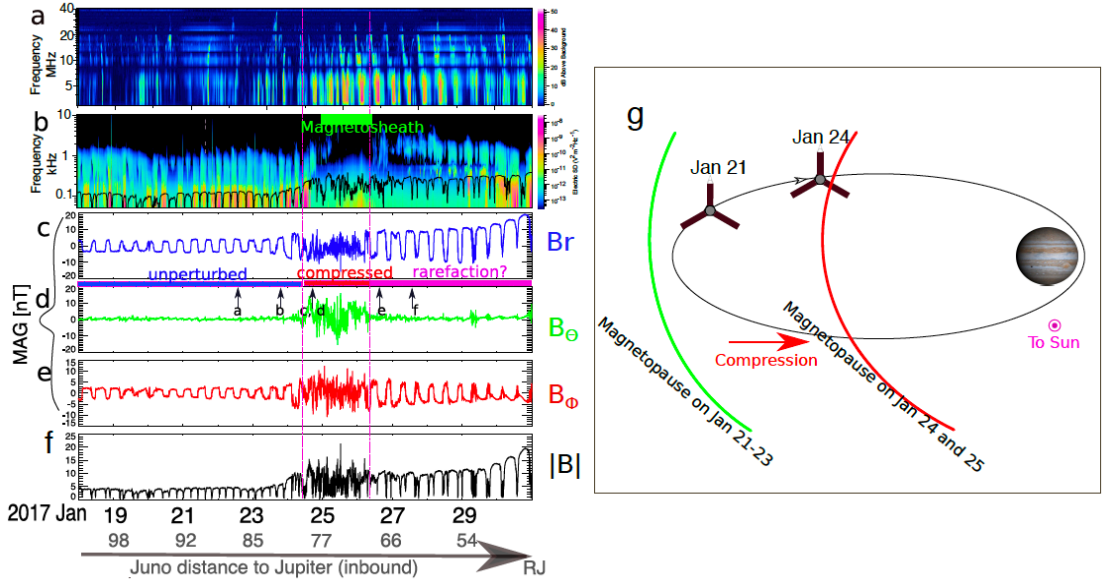

**Figure 2| Juno's measurements of waves and the component magnetic fields in System III coordinate system, showing unperturbed, strongly compressed and potentially expanding magnetosphere conditions.** a) Plasma wave spectrogram of hectometric and decametric emissions (a few to tens of MHz). b) Plasma wave spectrogram of electric field from 50 Hz to 10 keV. The disappearance and appearance of ~ 1 kHz continual emission indicate the entry and exit of Juno into the magnetosheath. c-f) Three components of magnetic fields and the magnetic strength. As marked on the top of panel d, we divide the observations into three periods, i.e., unperturbed, compressed and rarefaction conditions. g) A sketch to illustrate Juno's trajectory and the locations of magnetopause before and after compression. Note that the times for images in Fig. 1 are marked with black arrows in panel d. The electric field wave intensities were computed using the geometric antenna length of 2.4 meters.

**Table 1| The event list for ADS and MAB auroral morphologies.**

| Events [a] | Compression? If yes, encounter the magnetopause at when and where? | Juno location ($R_J$) | Arc Width /Variation (km) |
|---|---|---|---|
| ADS events | | | |
| 2016/07/18 18:58UT | Yes, 2016/07/17 00:09UT, at 91 $R_J$ | 96 | 2194/958 |
| 2017/01/22 15:31UT | No | 86 | 1468/548 |
| 2017/04/23 14:00UT | No | 113 | 1714/641 |
| MAB events [b] | | | |
| 2016/06/30 04:13UT | Yes, 2016/06/29 23:40UT, at 75 $R_J$ | 72 | 825/341 |
| 2016/07/14 16:23UT[c] | Yes, 2016/07/14 12:39UT, at 80 $R_J$ | 81 | 790/278 |
| 2016/07/17 14:21UT | Yes, 2016/07/17 00:09UT, at 91 $R_J$ | 92 | 830/289 |
| 2017/01/24 15:11UT[d] | Yes, 2017/01/24 17:30UT, at 78 $R_J$ | 78 | 626/262 |
| 2017/03/19 09:57UT[e] | Yes, inferred from modeling | 74 | 639/385 |

[a] The events were selected from 2016 June to 2017 July, when Juno was at >70 $R_J$ and simultaneous auroral images were available from HST.

[b] Note that the two MAB events on July 14 and 17 2016 may be grouped as a long-last solar wind compression event, but we could not confirm if the magnetopause or auroral morphology in between have returned to quiet condition.

[c] At ~2016/07/14 12:39UT, Juno encountered the magnetopause boundary layer, and clearly entered into the magnetosheath at 21:19 UT[34].

[d] At ~2017/01/24 17:30UT, Juno encountered the magnetopause boundary layer, and clearly entered into the magnetosheath at 2017/01/25 01:25UT[34].

[e] This auroral event and the solar wind compression condition are analyzed in details by Yao, et al. [35].

We surveyed the HST dataset from 2016 June to 2017 July when Juno was exploring the magnetosphere at > 70 $R_J$, to seek a systematic relation between the compressed magnetopause and the two types of auroral morphologies (i.e., MAB and ADS). Considering the nominal magnetopause boundary on the dawnside (i.e., the Juno trajectory) is about 110 $R_J$, we therefore empirically define the uncompressed magnetopause conditions by satisfying at least one of the two criteria: 1) Juno was at > 85 $R_J$ and did not observe substantial magnetic perturbations, and 2) Juno remains in the magnetosphere at > 110 $R_J$. The compressed magnetopause is defined when Juno directly encountered the magnetopause at < 90 $R_J$. As shown in Table 1, we have identified eight auroral events with coordinated Juno's in situ measurements and HST's remote sensing of aurorae (with an exception on March 19 event). Three of them are ADS morphology, and the other five are MAB morphology. As we introduced in the 2017 January 22-24 case study, the mean arc width and the variation parameters can be used to characterize each type of auroral morphology. We could therefore empirically identify ADS events with the mean arc width exceeding 1400 km and the variation exceeding 500 km. Similarly, if both the mean arc width is below 1000 km and the variation is below 400 km, we may empirically identify them as MAB events. The quiet aurora morphology may be empirically defined as total auroral power below 1200 GW, and the maximum brightness to be lower than 1000 kiloRayleighs on dawn side main auroral oval. It is noteworthy that these thresholds are empirical and based on a limited number of cases. A further statistical study using many more observations are important to refine the criteria. The enhanced solar wind compression was given by Tao model prediction[25] for the event on March 19 2017, and the auroral morphology is a typical MAB, consistent with the other five events, whose magnetopause compression were directly determined by plasma waves.

**Discussion**

Previous work has suggested that the main auroral oval also exhibits substantially reduced brightness near noon local time[36] (e.g., clearly shown in Fig. 1c and 1d). Traditionally, the auroral discontinuity is explained as a consequence of stronger solar wind compression, which reduces auroral precipitation in near noon local times. Consequently, the auroral discontinuity was suggested as evidence in supporting theoretical prediction[37]. However, the quantitative analysis (Fig. 3) of the auroral evolution rejects the hypothesis that the solar wind compressions cause the near noon aurora to dim.

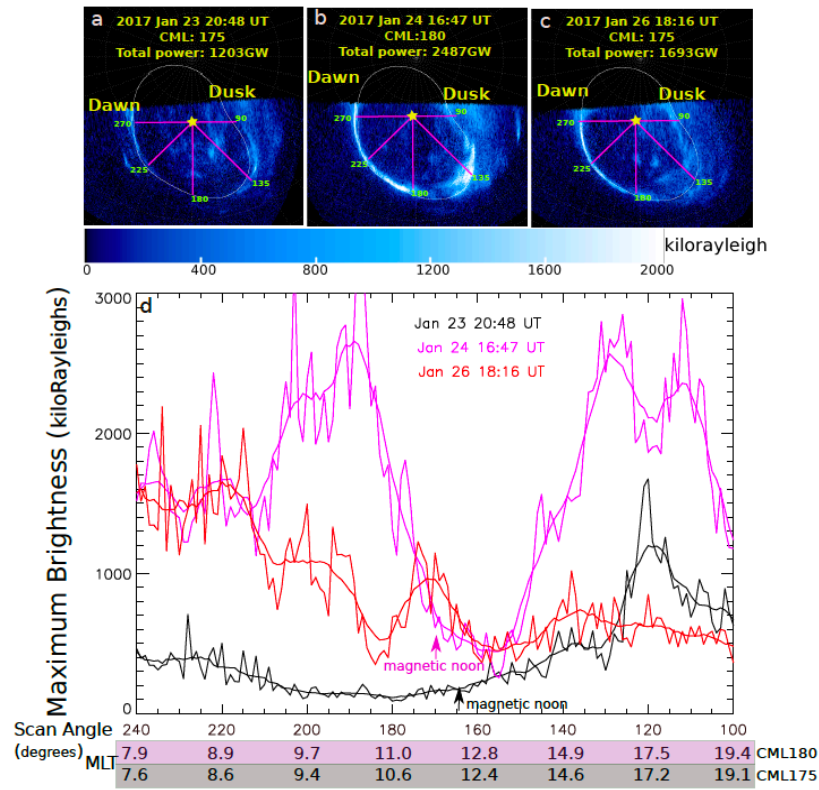

**Figure 3| Comparisons between auroral brightness distributions before, during and after magnetopause compression.** Top: three selected auroral images with similar viewing geometries. The yellow star indicates the morphological center of the main oval (System III longitude at 185° and latitude at 74°)[38], and the pink lines indicate scan angles at five given values. Bottom: distribution of maximum brightness as a function of scan angle for the three auroral images on the top panel during uncompressed magnetopause, compressed magnetopause and two days after the compression.

Fig. 3a, b and c show three selected images from the three HST visits (Fig. 1b, 1d and 1e), but each image is averaged over 1 minute. All the three images are from a similar viewing geometry, i.e., their Central Meridian Longitude (CML) numbers are similar (175, 180 and 175 respectively), which provide advantages to compare the local time and scan angle information at the same time (i.e., the X-axis). The magnetic local times obtained using flux equivalence model[39,40] while with JRM09[41] as an internal model, are overlaied along the scan angle. The slight difference in CML results in a difference of about 0.3 hour in MLT. As illustrated by the pink lines (original and smoothed over 10 points), the brightness in near noon local times (~500 kiloRayleighs at the scan angle of ~160 degrees) is much lower than dawn local times (higher than 2600 kilorayleighs at the scan angle of ~190 degrees) and dusk local

times (higher than 2600 kilorayleighs at the scan angle of ~130 degrees). The change is as large as 70 kiloRayleighs per degree. In contrast to the compressional period, the variation of auroral brightness along the main oval during quiet period is only ~10 kiloRayleighs per degree. Although solar wind compression enhanced the near noon auroral discontinuity (i.e., the gradient of auroral intensity) by a factor of 7, the observations do not support the hypothesis that solar wind compression dim near noon aurora. Oppositely, the auroral brightness in auroral discontinuity region also increased, although not by as much as the both sides of the discontinuity, which is why the discontinuity becomes clearly visible. The enhanced auroral discontinuity is near magnetic noon, and the quiet time auroral discontinuity is centered at about 10 MLT. The auroral distributions before and during the solar wind compression is generally consistent with the simulation results of field-aligned currents in Fig. 8 in Chané, et al. [18]. They show that the magnetic field lines are very different (in elongation) for different local time, and that this influences how much currents the corotation breakdown generates. The simulations cannot reproduce the exact position of the discontinuity (because the dipole tilt is not present), but the general behavior is well reproduced. This local time asymmetry is enhanced during solar wind compression periods, so that field-aligned currents are enhanced in all local times. The time-varying modeling results[17] predict that the magnetosphere would re-establish a steady state after 1-2 days of compression and the main aurora would be fainter than pre-compression state. This is, however, not supported by the auroral image shown in Fig. 3. This inconsistency was also revealed by Nichols, et al. [24].

Using the accurate determination of magnetopause compression and the contemporaneous auroral observation from HST, we reveal that the MAB auroral morphology is directly driven by solar wind compression, while ADS could occur during quiet and enhanced solar wind periods[24,29]. ADS events are substantially extended to lower latitudes, which may imply that energy sources for ADS span a large radial range from the middle to inner magnetosphere. Besides the large scale electrical current system like the current loop associated with the corotation breakdown enhancement force at Jupiter or substorm current wedge at Earth, electromagnetic waves (Alfvénic waves) are known to provide substantial contribution to global auroral intensifications[42]. The compression of magnetosphere is recently confirmed to produce intense Poynting flux and power aurora at Earth[43]. Since the MAB events are also related to solar wind compression, we thus suggest that solar wind compression produce MAB via Alfvénic fluctuations. Theoretical and observations studies have also confirmed the important roles of Alfvénic fluctuation in powering Jupiter's main aurora[44-46]. We notice that the hectometric radio emission was enhanced during all MAB events, but not during the two ADS events on January 22 2017 and April 23 2017, when the magnetopause was not compressed. The relationship between radio emission and UV auroral morphologies could provide insights in understanding the auroral driving mechanisms, although we also notice that the radio enhancement may last for longer time than UV aurora, which has also been reported in previous literature[13]. Further studies on their detailed relations are probably important to understand their systematic connections to solar wind compressions. The different involvements of solar wind compression in driving aurorae at Jupiter are analogous to terrestrial aurorae during geomagnetic storms and substorms. At Earth, magnetic storms are driven by external sources that reduce the magnetopause stand-off distance, while substorm expansions correspond to internal disruption of magnetotail currents with an unpredictable onset time[47,48], although the mass and energy originally come from solar wind or ionosphere. Note that the current disruption process could be a consequence of reconnection flow pileup[49,50] or near-Earth plasma instabilities[51].

**Methods: the determination of mean auroral width and its standard deviation**
Here we take Fig. 1a as an example to demonstrate how we calculate the mean auroral width and its standard deviation. In general, there are four major steps, as described below. Note that all the auroral images are taken from the northern hemisphere in this study.

**Step 1**. Define the scan angle system in polar projection
The auroral image (Fig. 4) on the grid (white dotted lines) is in System III coordinates[52], which corotates with the planet. The green and pink numbers indicate the System III longitudes and latitudes. The red star denotes the center of main auroral oval[36,38], also known as auroral oval's barycenter[53]. The yellow lines indicate scan angles, radiating from the auroral barycenter.

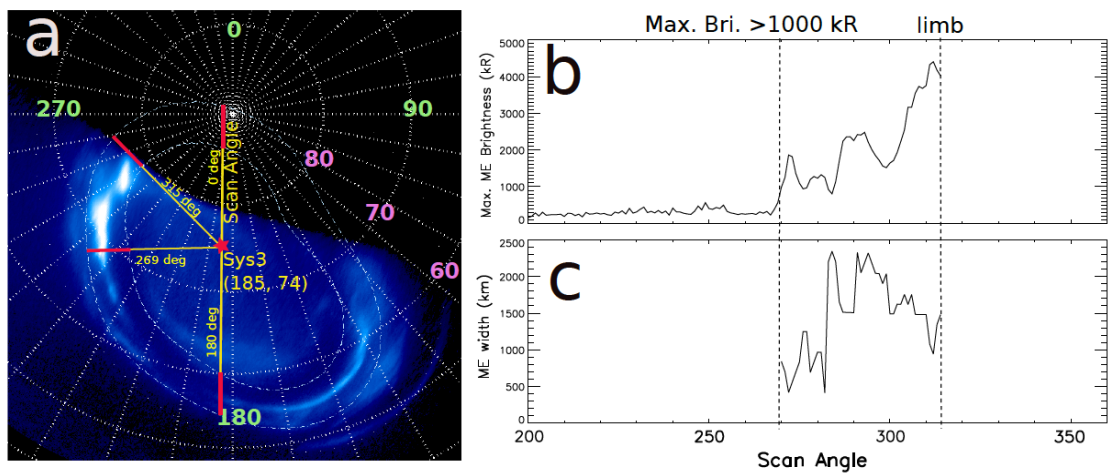

**Figure 4| Calculation of the main auroral arc width.** a) An auroral example from Fig. 1a. The yellow lines indicate scan angles, and the red star indicates auroral oval's center (System III longitude at 185° and latitude at 74°). b) Maximum brightness on the main emission along the scan angles. c) the width of main auroral arc along the scan angles.

**Step 2**. Identify the auroral maximum brightness along the main oval

We select the maximum brightness along each scan angle in a relatively large area, whose inner and outer boundaries are marked by the dash dot ovals on the auroral image. As indicated by the red bars in panel (a), it is clear that the region could well cover the main auroral emissions. The maximum brightness as a function of scan angle is shown in panel b.

**Step 3**. Calculate the width of the main emission ~perpendicular to the auroral oval for each scan angle
The width of main emission in panel c is obtained using the boundary of 50% of maximum brightness for each scan angle.

**Step 4**. Calculate mean arc width of the dawn aurora and the variation of thickness
As illustrated by the two vertical dashed lines in panels (b and c), we determine the range of scan angle to calculate the auroral mean width and its standard deviation based on two features, 1) the initial scan angle is determined by intensities > 1000 kiloRayleighs, 2) the final scan angle is determined by the upper limit of scan angle imparted by the viewing geometry in the polar projection. After determining the scan angle range and width at each scan angle, we thus directly calculate the mean arc width of the dawn aurora and its standard deviation as the variation of arc thickness.

**Acknowledgements:**
Z.Y. acknowledges the Strategic Priority Research Program of Chinese Academy of Sciences (Grant No. XDA17010201). B.B. is a Research Associate of the Fonds de la Recherche Scientifique - FNRS. We are grateful to NASA and contributing institutions which have made the Juno mission possible. This work was funded by NASA's New Frontiers Program for Juno via contract with the Southwest Research Institute. B.B., D.G., B.P. and R.L.G. acknowledge financial support from the Belgian Federal Science Policy Office (BELSPO) via the PRODEX Programme of ESA. The research at the University of Iowa was supported by NASA through Contract 699041X with Southwest Research Institute.

**Data Availability**

The auroral images are based on observations with the NASA/ESA Hubble Space Telescope (program HST GO-14105 and GO-14634), obtained at the Space Telescope Science Institute (STScI), which is operated by AURA for NASA. All data are publicly available at STScI via https://archive.stsci.edu/hst/. All Juno data presented here are publicly available from NASA's Planetary Data System (https://pds-ppi.igpp.ucla.edu/) as part of the JNO-J-3-FGM-CAL-V1.0, and JNO-E/J/SS-WAV-2-EDR-V1.0 datasets for the MAG and Wave instruments.

**References:**

1. Mauk, B. & Bagenal, F. Comparative auroral physics: Earth and other planets. *Auroral Phenomenology and Magnetospheric Processes: Earth and Other Planets*, 3-26 (2013).
2. Waite Jr, J. *et al.* An auroral flare at Jupiter. *Nature* **410**, 787 (2001).
3. Kimura, T. *et al.* Transient internally driven aurora at Jupiter discovered by Hisaki and the Hubble Space Telescope. *Geophysical Research Letters* **42**, 1662-1668 (2015).
4. Connerney, J. & Satoh, T. The H3+ ion: A remote diagnostic of the Jovian magnetosphere. *Philosophical Transactions of the Royal Society of London. Series A: Mathematical, Physical and Engineering Sciences* **358**, 2471-2483 (2000).
5. Dunn, W. *et al.* The independent pulsations of Jupiter's northern and southern X-ray auroras. *Nature Astronomy* **1**, 758 (2017).
6. Gladstone, G. *et al.* A pulsating auroral X-ray hot spot on Jupiter. *Nature* **415**, 1000-1003 (2002).
7. Kurth, W., Barbosa, D., Scarf, F., Gurnett, D. & Poynter, R. Low frequency radio emissions from Jupiter: Jovian kilometric radiation. *Geophysical Research Letters* **6**, 747-750 (1979).
8. Grodent, D. *et al.* Jupiter's polar auroral emissions. *Journal of Geophysical Research: Space Physics* **108** (2003).
9. Cowley, S. & Bunce, E. Origin of the main auroral oval in Jupiter's coupled magnetosphere–ionosphere system. *Planetary and Space Science* **49**, 1067-1088 (2001).
10. Hill, T. The Jovian auroral oval. *Journal of Geophysical Research: Space Physics* **106**, 8101-8107 (2001).
11. Southwood, D. & Kivelson, M. A new perspective concerning the influence of the solar wind on the Jovian magnetosphere. *Journal of Geophysical Research: Space Physics* **106**, 6123-6130 (2001).
12. Baron, R., Owen, T., Connerney, J., Satoh, T. & Harrington, J. Solar wind control of Jupiter's H+ 3auroras. *Icarus* **120**, 437-442 (1996).
13. Gurnett, D. *et al.* Control of Jupiter's radio emission and aurorae by the solar wind. *Nature* **415**, 985 (2002).
14. Nichols, J. *et al.* Response of Jupiter's UV auroras to interplanetary conditions as observed by the Hubble Space Telescope during the Cassini flyby campaign. *Journal of Geophysical Research: Space Physics* **112** (2007).
15. Sinclair, J. *et al.* A brightening of Jupiter's auroral 7.8-μm CH 4 emission during a solar-wind compression. *Nature Astronomy*, 1 (2019).
16. Hill, T. Inertial limit on corotation. *Journal of Geophysical Research: Space Physics* **84**, 6554-6558 (1979).
17. Cowley, S., Nichols, J. & Andrews, D. J. Modulation of Jupiter's plasma flow,

18  Chané, E., Saur, J., Keppens, R. & Poedts, S. How is the Jovian main auroral emission affected by the solar wind? *Journal of Geophysical Research: Space Physics* **122**, 1960-1978, doi:doi:10.1002/2016JA023318 (2017).

19  Kita, H. *et al.* Jovian UV aurora's response to the solar wind: Hisaki EXCEED and Juno observations. *Journal of Geophysical Research: Space Physics* (2019).

20  Hospodarsky, G. *et al.* Jovian bow shock and magnetopause encounters by the Juno spacecraft. *Geophysical Research Letters* **44**, 4506-4512 (2017).

21  Grodent, D. *et al.* Jupiter's aurora observed with HST during Juno orbits 3 to 7. *Journal of Geophysical Research: Space Physics*, doi:10.1002/2017JA025046 (2018).

22  Clarke, J. T. *et al.* Hubble Space Telescope imaging of Jupiter's UV aurora during the Galileo orbiter mission. *Journal of Geophysical Research: Planets* **103**, 20217-20236 (1998).

23  Kita, H. *et al.* Characteristics of solar wind control on Jovian UV auroral activity deciphered by long‐term Hisaki EXCEED observations: Evidence of preconditioning of the magnetosphere? *Geophysical Research Letters* **43**, 6790-6798 (2016).

24  Nichols, J. *et al.* Response of Jupiter's auroras to conditions in the interplanetary medium as measured by the Hubble Space Telescope and Juno. *Geophysical Research Letters* **44**, 7643-7652, doi:10.1002/2017GL073029 (2017).

25  Tao, C., Kataoka, R., Fukunishi, H., Takahashi, Y. & Yokoyama, T. Magnetic field variations in the Jovian magnetotail induced by solar wind dynamic pressure enhancements. *Journal of Geophysical Research: Space Physics* **110** (2005).

26  Kurth, W. *et al.* The Juno waves investigation. *Space Science Reviews* **213**, 347-392 (2017).

27  Connerney, J. *et al.* The Juno magnetic field investigation. *Space Science Reviews* **213**, 39-138 (2017).

28  Joy, S. *et al.* Probabilistic models of the Jovian magnetopause and bow shock locations. *Journal of Geophysical Research: Space Physics* **107** (2002).

29  Kimura, T. *et al.* Transient brightening of Jupiter's aurora observed by the Hisaki satellite and Hubble Space Telescope during approach phase of the Juno spacecraft. *Geophysical Research Letters* **44**, 4523-4531 (2017).

30  Gurnett, D., Kurth, W. & Scarf, F. The structure of the Jovian magnetotail from plasma wave observations. *Geophysical Research Letters* **7**, 53-56 (1980).

31  Scarf, F. L., Gurnett, D. A. & Kurth, W. S. Jupiter plasma wave observations:

An initial Voyager 1 overview. *Science* **204**, 991-995 (1979).

32  Kurth, W. *et al.* The dusk flank of Jupiter's magnetosphere. *Nature* **415**, 991 (2002).

33  Gershman, D. J. *et al.* Juno observations of large‐scale compressions of Jupiter's dawnside magnetopause. *Geophysical Research Letters* **44**, 7559-7568 (2017).

34  Ranquist, D. *et al.* Survey of Jupiter's Dawn Magnetosheath Using Juno. *Journal of Geophysical Research: Space Physics* **124**, 9106-9123 (2019).

35  Yao, Z. *et al.* On the relation between Jovian aurorae and the loading/unloading of the magnetic flux: simultaneous measurements from Juno, HST and Hisaki. *Geophys. Res. Lett* **46**, doi:10.1029/2019GL084201 (2019).

36  Radioti, A. *et al.* Discontinuity in Jupiter's main auroral oval. *Journal of Geophysical Research: Space Physics* **113** (2008).

37  Cowley, S. *et al.* A simple axisymmetric model of magnetosphere‐ionosphere coupling currents in Jupiter's polar ionosphere. *Journal of Geophysical Research: Space Physics* **110** (2005).

38  Grodent, D., Gérard, J. C., Clarke, J., Gladstone, G. & Waite, J. A possible auroral signature of a magnetotail reconnection process on Jupiter. *Journal of Geophysical Research: Space Physics* **109** (2004).

39  Vogt, M. F. *et al.* Magnetosphere‐ionosphere mapping at Jupiter: Quantifying the effects of using different internal field models. *Journal of Geophysical Research: Space Physics* **120**, 2584-2599 (2015).

40  Vogt, M. F. *et al.* Improved mapping of Jupiter's auroral features to magnetospheric sources. *Journal of Geophysical Research: Space Physics* **116** (2011).

41  Connerney, J. *et al.* A new model of Jupiter's magnetic field from Juno's first nine orbits. *Geophysical Research Letters* **45**, 2590-2596 (2018).

42  Keiling, A., Wygant, J., Cattell, C., Mozer, F. & Russell, C. The global morphology of wave Poynting flux: Powering the aurora. *Science* **299**, 383-386 (2003).

43  Keiling, A., Thaller, S., Wygant, J. & Dombeck, J. Assessing the global Alfvén wave power flow into and out of the auroral acceleration region during geomagnetic storms. *Science Advances* **5**, eaav8411 (2019).

44  Gershman, D. J. *et al.* Alfvénic fluctuations associated with Jupiter's auroral emissions. *Geophysical Research Letters* **46**, 7157-7165 (2019).

45  Saur, J. *et al.* Wave‐Particle Interaction of Alfvén Waves in Jupiter's Magnetosphere: Auroral and Magnetospheric Particle Acceleration. *Journal of Geophysical Research: Space Physics* **123**, 9560-9573 (2018).

46  Mauk, B. *et al.* Discrete and broadband electron acceleration in Jupiter's powerful aurora. *Nature* **549**, 66 (2017).

47  Rostoker, G., Akasofu, S., Baumjohann, W., Kamide, Y. & McPherron, R. The roles of direct input of energy from the solar wind and unloading of stored magnetotail energy in driving magnetospheric substorms. *Space science reviews* **46**, 93-111 (1988).

48  Lewis, Z. On the apparent randomness of substorm onset. *Geophysical Research Letters* **18**, 1627-1630 (1991).

49  Angelopoulos, V. *et al.* Tail reconnection triggering substorm onset. *Science* **321**, 931-935, doi:10.1126/science.1160495 (2008).

50  Shiokawa, K., Baumjohann, W. & Haerendel, G. Braking of high-speed flows in the near-Earth tail. *Geophysical Research Letters* **24**, 1179-1182, doi:10.1029/97gl01062 (1997).

51  Lui, A. T. Y. Current disruption in the Earth's magnetosphere: Observations and models. *J Geophys Res-Space* **101**, 13067-13088, doi:10.1029/96ja00079 (1996).

52  Fränz, M. & Harper, D. Heliospheric coordinate systems. *Planetary and Space Science* **50**, 217-233 (2002).

53  Bonfond, B. *et al.* in *Ann Geophys-Germany.*  1203-1209 (2015).